\documentclass[aps,reprint,superscriptaddress,nofootinbib,showpacs]{revtex4-1}
\usepackage{amsmath,latexsym,amssymb,hyperref,graphicx}
\usepackage{slashed,nicefrac}

\hypersetup{colorlinks,citecolor= nicegreen,linkcolor= nicered}
\usepackage{color}
\definecolor{nicered}{rgb}{0.7,0.1,0.1}
\definecolor{nicegreen}{rgb}{0.1,0.5,0.1}
 
\newcommand\eV{\text{eV}}

\newcommand\MeV{\text{MeV}}
\newcommand\GeV{\text{GeV}}
\newcommand\TeV{\text{TeV}}
\newcommand\meV{\text{meV}}

\newcommand\BB{\ensuremath{0\nu2\beta}}

\newcommand\SEC[1]{\medskip\noindent{\sl\bfseries #1}}
\renewcommand{\figurename}{\renewcommand\baselinestretch{1.15}\scriptsize\em Fig.{}}
\renewcommand{\tablename}{\renewcommand\baselinestretch{1.15}\scriptsize\em  Table}

\begin{document}
\addtolength{\belowdisplayskip}{-.5ex}       
\addtolength{\abovedisplayskip}{-.5ex}       

\title{Left-Right Symmetry: from LHC to Neutrinoless Double Beta Decay}

\author{Vladimir Tello}
\affiliation{SISSA, Trieste, Italy}

\author{Miha Nemev\v{s}ek}
\affiliation{ICTP, Trieste, Italy}
\affiliation{Jo\v zef Stefan Institute, Ljubljana, Slovenia}

\author{Fabrizio Nesti}
\affiliation{Universit\`a di Ferrara, Ferrara, Italy}

\author{Goran Senjanovi\'{c}}
\affiliation{ICTP, Trieste, Italy}

\author{Francesco Vissani}
\affiliation{LNGS, INFN, Assergi, Italy}

\date{\today}

\begin{abstract}
  \noindent 
  The Large Hadron Collider has a potential to probe the scale of left-right symmetry restoration and the
  associated lepton number violation.  Moreover, it offers hope of measuring the right-handed leptonic mixing
  matrix. We show how this, together with constraints from lepton flavor violating processes, can be used to
  make predictions for neutrinoless double beta decay.  We illustrate this connection in the case of the
  type-II seesaw.
\end{abstract}


\maketitle

\noindent 
More than 70 years ago Majorana \cite{Majorana:1937vz} raised the question whether neutrinos are ``real''
particles. If true, this
would allow for neutrinoless double beta decay ($0\nu2\beta$)~\cite{racahfurry}, a violation of lepton number with two
electrons created out of ``nothing''. The transition amplitude is proportional to
\begin{equation}
	\mathcal A_\nu \propto {\small G_F^2} \frac{\displaystyle m_\nu^{ee}}{\displaystyle p^2},
\end{equation}
where $m_\nu^{ee}$ is the 1-1 element of the neutrino mass matrix $m_\nu$ and $p \approx 100\,\MeV$
a measure of the neutrino virtuality. Present-day neutrinoless double beta experiments are probing
the sub-eV region for $m_\nu^{ee}$. There is even a claim of this process being seen, corresponding
to $m_\nu^{ee} \approx 0.4\,\eV$ \cite{KlapdorKleingrothaus:2004wj}.  On the other hand, the upper
limits on the sum of neutrino masses from cosmology are rapidly progressing and recently, it was
argued that the two are incompatible~\cite{Fogli:2008ig}.  Whether or not such a conclusion is
premature today, we should consider seriously the possibility that this minimal picture will be
contradicted by the next round of experiments~\cite{cuoregerda}.

This would imply new physics doing the job~\cite{feinbergbrown}, whose contribution to the transition amplitude can be cast in the natural form
\vspace*{-2ex}
\begin{equation}
	\mathcal A_{\text{NP}} \propto G_F^2 \frac{\displaystyle M_W^4}{\displaystyle \Lambda^5},
\end{equation}
where $\Lambda$ is the scale of new physics. The new physics enters the game at $\Lambda \sim\TeV$,
tailor-made for the Large Hadron Collider (LHC), which provides a strong
motivation to pursue this line of thought.

A natural candidate for new physics is the right-handed charged current, as argued~\cite{ms} 
in the context of left-right (LR) symmetric theories~\cite{lr}. It was
precisely LR symmetry that led to neutrino masses and, on
top, connected them~\cite{minkowskims} to the scale of parity restoration in the context of the
seesaw mechanism~\cite {seesaw, minkowskims}.  This leads to a
remarkable signature of lepton number violation\pagebreak[3] in the form of same sign lepton pairs
at colliders~\cite{Keung:1983uu} in complete analogy with \BB.
Furthermore, with such a low scale one expects sizable rates for lepton flavor violating
(LFV) processes, which are being vigorously pursued in the ongoing and planned experiments, yet
another encouragement to follow the road of new physics.

\begin{figure}
	\includegraphics[width=4.1cm]{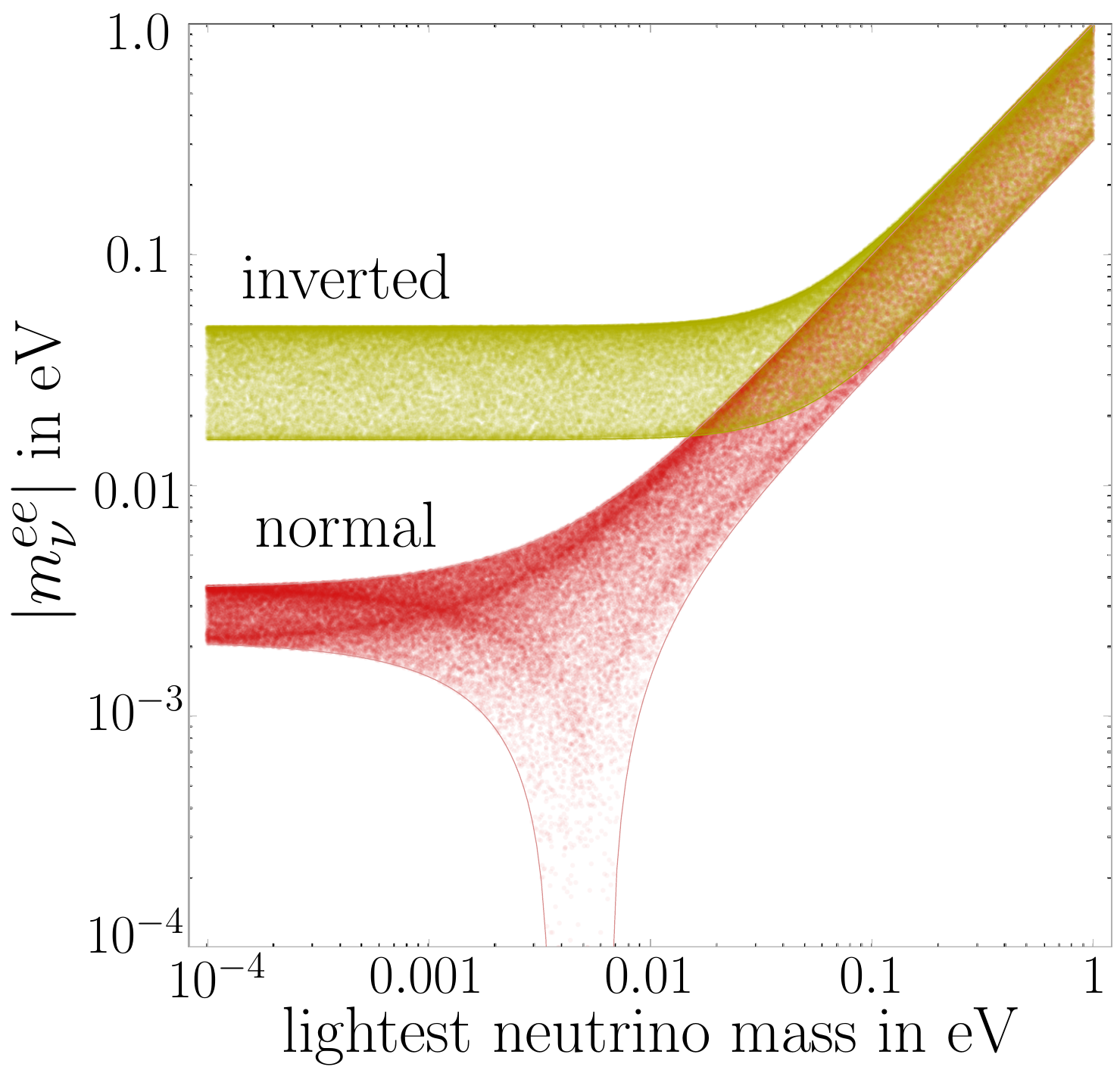}
	\includegraphics[width=4.1cm]{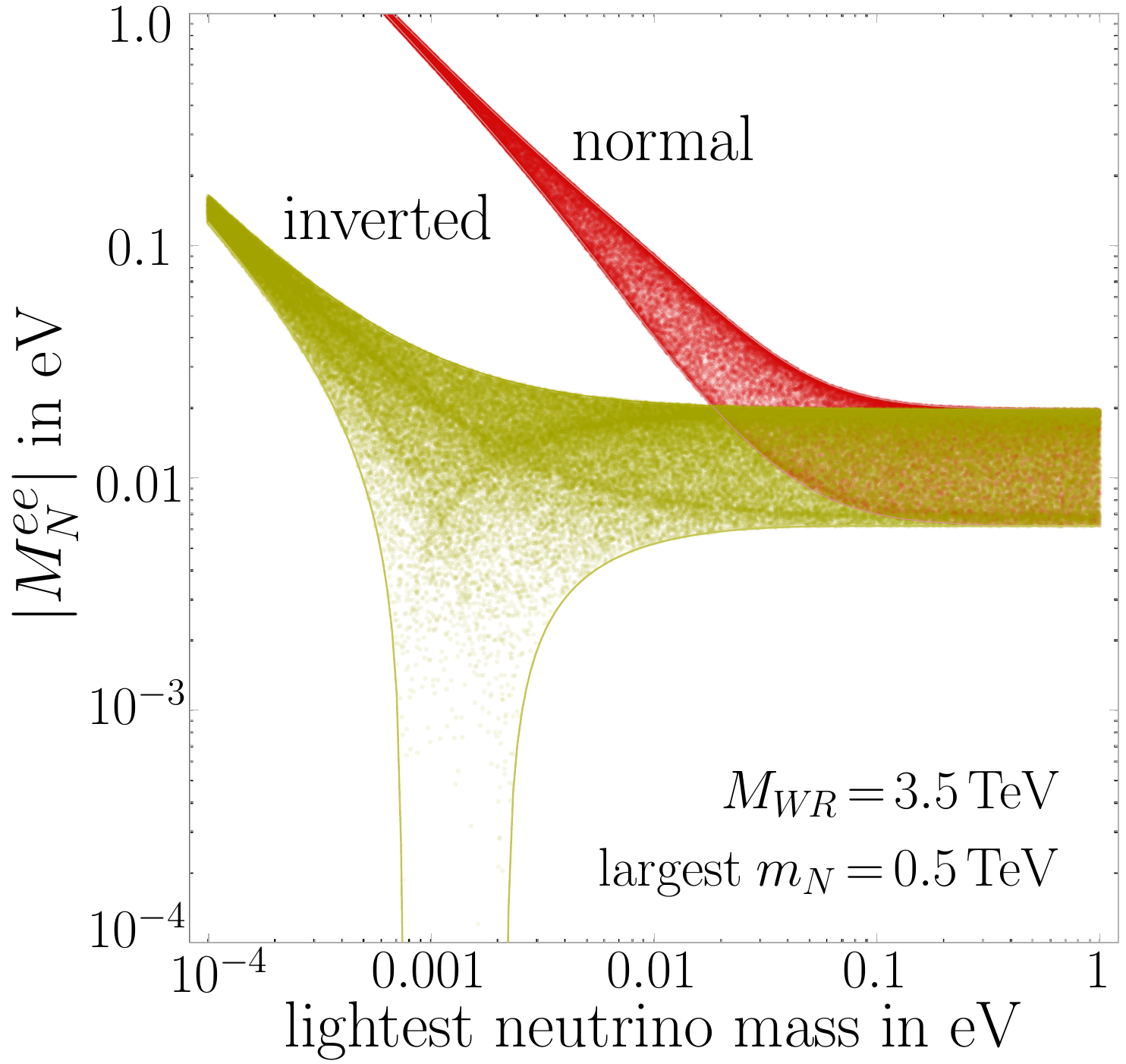}%
        \vspace*{-1ex}%
	\caption{The canonical contribution (left) from light neutrino mass and the new physics
          part (right), with $|M_{\scriptscriptstyle {N} }^{ee}|$ defined in Eq.~\eqref{eqMNee}.
          The mixing angles are fixed at $\{\theta_{12}, \theta_{23}, \theta_{13}\} = \{35^{\circ}
          ,45^{\circ},7^{\circ}\}$, while the Dirac and Majorana phases vary in the interval
          $[0,2\pi]$.}
        \vspace{-2ex}
        \label{figTelloPlot}%
\end{figure}

Motivated by these considerations, we have performed a detailed study of the relation between LHC, $0\nu
2\beta$ and LFV, in the context of the minimal LR model with type II seesaw. Our main point is shown in
Fig.~\ref{figTelloPlot}, where the new physics contribution is contrasted with the usual one, due to neutrino
mass \cite{Vissani:1999tu}. Since the standard contribution entails $m_\nu^{ee}$, we use a combination of new physics parameters
with the same dimension, denoted hereafter as $M_N^{ee}$. It depends on the mass of the right-handed charged
gauge boson and on masses and mixings of the heavy right-handed neutrinos as displayed below in
Eq.~\eqref{eqMNee}.

The striking feature which emerges is the reversed role of neutrino mass hierarchies. While in the case of
neutrino mass behind neutrinoless double beta decay the normal hierarchy matters less and degeneracy is most
promising, in the case of new physics it is normal hierarchy that dominates and degeneracy matters less.  This
conclusion is true when the scale of new physics lies within the LHC reach~\cite{Ferrari:2000sp}. In other
words, the discovery of LR symmetry at LHC would provide an additional boost for neutrinoless double beta
decay searches. This is the main message of our Letter. In the following we describe the model and analyze its
predictions.

\SEC{The Model.} The minimal LR symmetric theory is based on the gauge group $\mathcal G_{LR} = SU(2)_L \times
SU(2)_R \times U(1)_{B-L}$ and a symmetry between the left and right sectors~\cite{lr},
which can be taken to be charge conjugation $\mathcal C$ (for the advantages of
this choice, see~\cite{Maiezza:2010ic}). Fermions are LR symmetric, $q_{L,R }= (u, d)_{L,R}$ and $\ell_{L,R} =
\left( \nu, e \right)_{L,R}$, with $f_{L} \leftrightarrow (f_{R})^c$ under $\mathcal C$, and the gauge
couplings are $g_L=g_R\equiv g$.

The Higgs sector consists~\cite{minkowskims} of the $SU(2)_{L,R}$
triplets\footnote{There is also a bidoublet, which takes the usual role of the SM Higgs doublet, and we
  do not discuss it here.  For a recent detailed analysis of its phenomenology and limits on its
  spectrum, see~\cite {Maiezza:2010ic}.} 
  $\Delta_{L, R} = \left(\Delta^{++}, \Delta^+, \Delta^0 \right)_{L, R}$,
  $\Delta_L \in (3 , 1 ,2)$ and $\Delta_R \in (1, 3, 2)$, 
  which under $\mathcal C$
transform as $\Delta_L \leftrightarrow \Delta_R^*$. The group $\mathcal G_{LR}$ is broken down to
the Standard Model (SM) gauge group by $\langle \Delta_R \rangle \gg M_W$ and after the SM symmetry
breaking, the left-handed triplet develops a tiny $\langle \Delta_L \rangle \ll M_W$. $\langle \Delta_R \rangle$ gives masses not only to the $W_R$ and $Z_R$ gauge bosons but also
to the right-handed neutrinos. 

The symmetric Yukawa couplings of the triplets relevant for our discussion are
\begin{equation}
	{\cal L}_Y  = \frac{1}{2} \ell_L  \frac{M_{\nu_L}}{\langle \Delta_L\rangle}
	 \Delta_L \ell_L +
	\frac{1}{2} \ell_R  \frac{M_{\nu_R}}{\langle \Delta_R\rangle}  \Delta_R \ell_R+ 
\text{h.c.}\,,
\label{eqLDelta}
\end{equation}
where $M_{\nu_L}$ and $M_{\nu_R}$ are Majorana mass matrices of light and heavy neutrinos. In principle, there
are also Dirac Yukawa couplings connecting the two. When these tiny couplings play a negligible role, the
resulting seesaw is called type II~\cite{typeII}. Purely for reasons of illustration, the rest of this Letter will be devoted to this appealing case. Due to $\mathcal C$, its
main characteristic is the connection between the two neutrino mass matrices $M_{\nu_R}/\langle \Delta_R
\rangle = M_{\nu_L}^*/\langle \Delta_L \rangle^*$.  An immediate 
consequence 
is the proportionality
of the two mass spectra
\begin{equation}
	m_N \propto m_\nu\,,
\label{spectrum}
\end{equation}
where $m_N$ stands for the masses of the three heavy right-handed neutrinos $N_i$ and $m_\nu$ for those of the
three light left-handed neutrinos $\nu_i$. 

In this theory, there are both left and right-handed charged gauge bosons with their corresponding leptonic
interactions in the mass eigenstate basis:
\begin{equation}
\mathcal L_{W} =  \frac{g}{\sqrt 2} \left(
\bar \nu_L V_{L}^\dag \slashed{W}\!_L e_L +
\bar N_R V_{R}^\dag \slashed{W}\!_R e_R\right)
+\text{h.c.}\,.
\end{equation}
Since the charged fermion mass matrices are symmetric (due to the symmetry under $\mathcal C$), one readily
obtains a connection (up to complex phases, irrelevant to our discussion) between the right-handed and the
left-handed (PMNS) leptonic mixings matrices
\begin{equation}
	V_R = V_L^*.
	\label{rightmixing}
\end{equation}

\SEC{LHC signatures or How to check type II.} LHC offers an exciting possibility of seeing directly both LR
symmetry restoration and lepton number violation. The point is that once produced, $W_R$ can decay into a
charged lepton and a right-handed neutrino which in turn decays into a second charged lepton and two jets.
Being Majorana particles, they decay into both leptons and anti-leptons, hence one obtains same sign lepton
pairs, signaling the violation of lepton number~\cite{Keung:1983uu}. It turns out that in this way, LHC
running at 14 TeV can reach $M_{W_R} \lesssim 2.1(4)\,\TeV$ with a luminosity of $0.1(30)\,
\text{fb}^{-1}$~\cite{Ferrari:2000sp}.  Since in the minimal model there is a rough bound of about $M_{W_R}
\gtrsim 2.5 \,\TeV$~\cite{Maiezza:2010ic}, in order to be conservative in our analysis we choose a
representative point $M_{W_R} = 3.5 \,\TeV$ together with $m_N^\text{heaviest} = 0.5 \,\TeV$.

The flavor dependence of $V_R$ can be determined precisely through these same sign lepton pair
channels; thus, Eq.~\eqref{rightmixing} may be falsified in the near future. 
Furthermore, if LHC will
measure the heavy right-handed masses in the same process, one could perform crucial consistency
checks of type II seesaw, such as
\begin{equation}\label{oscdata}
\frac{m_{N _2}^2-m_{N_1}^2}{m_{N _3}^2-m_{N _1}^2}=
\frac{m_{\nu _2}^2-m_{\nu_1}^2}{m_{\nu_3}^2-m_{\nu_1}^2} \simeq \pm 0.03\,.
\end{equation}
Here, the right-hand side is determined by oscillation data and the $\pm $ signs corresponds to
normal/inverted hierarchy case.  Another eloquent relation among the mass scales probed in cosmology,
atmospheric neutrino oscillations and LHC is:
\begin{equation}\label{eqGoldenFormula}
  m_{\text{cosm}}\! =\! 
  \sum_i m_{\nu_i} \simeq
  50\,\meV \times\! 
  \frac{\sum_i m_{N_i} }{\displaystyle \sqrt{|m_{N_3}^2 - m_{N_2}^2|}}.
\end{equation}
%

The bottom line is that the LHC can determine the right-handed neutrino masses and mixings and allow
one to make predictions studied below. The type II seesaw chosen here is only a transparent illustration
of how these 
connections take place.

\SEC{Lepton Flavor Violation.} Lepton flavor violation in LR symmetric theories has been studied in the
past~\cite{Cirigliano:2004mv}. What is new in our analysis is the connection with LHC and especially the
quantitative implications for \BB.
 
\begin{figure}
	\includegraphics[width= 4.1cm]{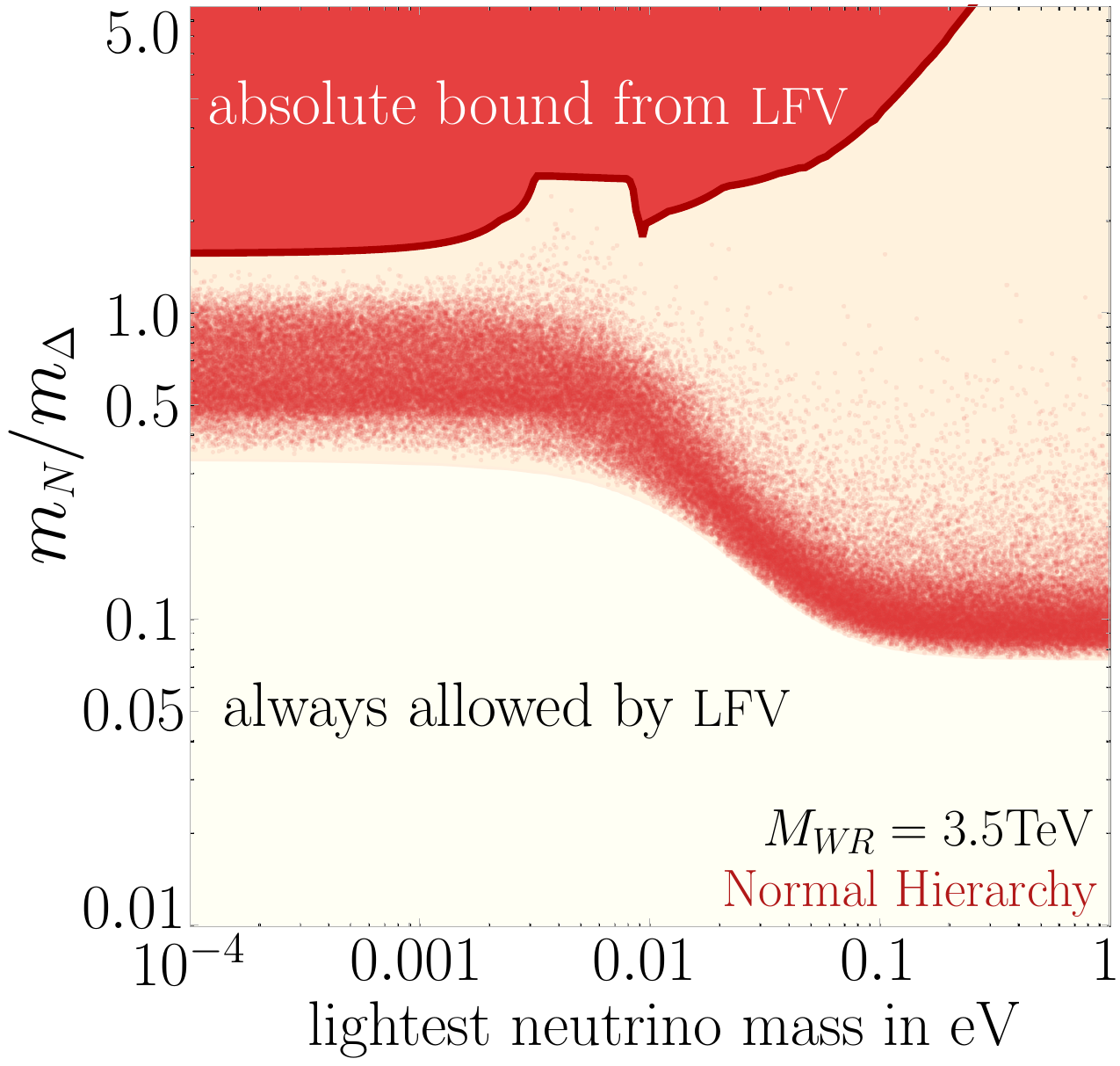}
	\includegraphics[width= 4.1cm]{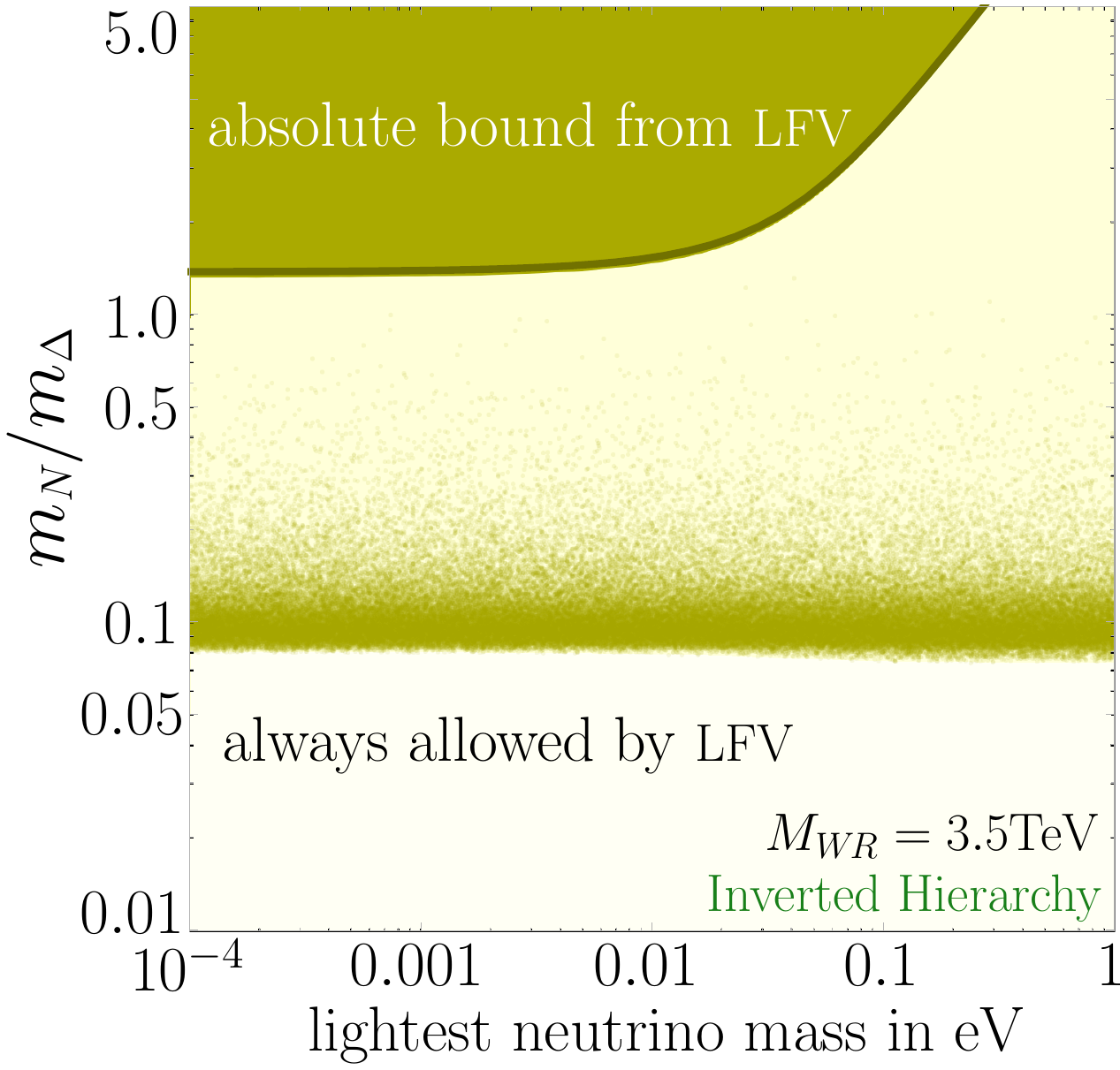}%
        \vspace*{-1ex}%
	\caption{\it Combined bounds on $m_N^\text{heaviest}/m_{\Delta}$ from LFV.  The dots show the (most
          probable) upper bounds resulting for different mixing angles and phases (varied respectively in the
          intervals $\{\theta_{12}, \theta_ {23}, \theta_{13}\}=$ $\{31^{\circ} \text{-}39^{\circ} ,37^{\circ}
          \text{-}53^{\circ}, 0 \text{-}13^{\circ}\}$ and $[0,2\pi]$).  The dark line is the
          absolute upper bound. The plot scales as $M_{W\hspace{-0.05cm}R}/3.5\,\TeV$.}
        \vspace{-0.1cm}%
        \label{figLFV} 
\end{figure}

There are various LFV processes providing constraints on the masses of right-handed neutrinos and doubly
charged scalars illustrated in Fig.~\ref{figLFV}.  It turns out that $\mu \rightarrow 3 e$, induced by the
doubly charged bosons $\Delta_{L}^{++}$ and $\Delta_{R}^{++}$, provides the most relevant constraint and so we
give the corresponding branching ratio
\begin{equation}
\!\!\!\!{\text{BR}}_{\mu \rightarrow 3 e} = \frac{1}{2}\!  \left(\!\frac{M_W}{M_{W_R}}\!  \right)^{\!\!4} \left|V_L \frac{m_N}
{m_{\Delta}}V_L^T \right|^2_{e \mu} \left|V_L \frac{m_N}{m_{\Delta }}V_L^T \right|^2_{ee},
\end{equation}
where $1/m_{\Delta}^2\equiv1/m_{\Delta_L}^2+1/m_{\Delta_R}^2$. The current experimental limit is $\text{ 
BR}(\mu\rightarrow 3e)<1.0 \times 10^{-12}$~\cite{Bellgardt:1987du}. 

The LFV transition rates become negligible when the masses $M_{W_R}$ and $m_{\Delta}$ become larger than about
100\,\TeV.  We are interested in LHC accessible energies, in which case the smallness of the LFV is governed
by the ratio $m_N/m_\Delta$, in addition to mixing angles and phases.  In Fig.~\ref{figLFV}, we plot the upper
bound on this quantity varying the mixing angles and phases (LFV plots also take into account $\mu\rightarrow
e$ conversion in Au nuclei, $\mu \rightarrow e \gamma$ and rare $\tau$ decays such as $\tau\rightarrow 3\mu$,
etc.~\cite{LFV}).  An immediate rough consequence seems to follow: $m_N^{\text{heaviest}} / m_{\Delta} < 0.1$
in most of the parameter space. However, the strong dependence on angles and phases allows this mass ratio up
to about one in the case of hierarchical neutrino spectra, thus allowing both $N$ and $\Delta_{L,R}$ to be
light.  This serves as an additional test at colliders of type II seesaw used here. For degenerate neutrinos,
unfortunately, no strict contraint arises: see again Fig.~\ref{figLFV}.

\SEC{Neutrinoless double beta decay.} We neglect the small neutrino Dirac Yukawa couplings, the tiny $W_L$-$W_R$ mixing of
$\mathcal{O}(M_W/M_{W_R})^2 \lesssim 10^{-3}$ and contributions coming from the bidoublet through
the charged Higgs, because of its heavy mass of at least 10\,\TeV~\cite{Maiezza:2010ic}. We
are 
left with an effective Hamiltonian with two extra contributions (the
one from the left-handed triplet being completely negligible)
\begin{equation} \label{eq:0nubblag}
\!\!\!\mathcal{H}_{\text{NP}}= G_F^2 V_{Rej}^2 \! \left[ \frac{1}{m_{N_j}}\!  +\! 
\frac{2 \ m_{N_j} }{ m^{2}_{\Delta_R^{++}} }  \right]\! \frac{M_W^4}
{M_{W_R}^4} \! J_{R\mu }^{\,}J^{\mu}_R\, \overline{e_{R}} e^{\,\,c}_R \,,
\end{equation}
where $J_{R \mu}$ is the right-handed hadronic current. Making use of the LFV constraint  
$m_N/m_{\Delta} \ll 1$ one can neglect the $\Delta_R^{++}$ contribution and write the total decay rate as
\begin{equation}
  \frac{\Gamma_{  0\nu\beta\beta }}{\text{ln\,2}} = G \cdot \left| \frac{\mathcal{M}_{\nu}}{m_e}\right|^2   \Bigg( |  
  m_{\nu}^{ee} |^2 + \Bigg| p^2 \frac{M_W^4}{M_{W_R}^4}  \frac{V_{Rej}^2}{m_{N_j}}   \Bigg|^2 \Bigg)\,,
\end{equation}
where $G$ is a phase space factor, $\mathcal{M}_{\nu}$ is the nuclear matrix element relevant for
the light neutrino exchange, while $p$ measures the neutrino virtuality and accounts also for the
ratio of matrix elements of heavy and light neutrinos. These quantities have been calculated and 
\cite{Hirsch:1996qw,Simkovic:2010ka} 
are reported in
Table~\ref{tab:nvirtuality} for some interesting nuclei.

 \begin{table}[t] 
  \vspace{-0.3cm}%
\footnotesize
$$
\begin{array}{c| c|c |c|c|c|c|c  }
\!\!\text{ref.} &\text{nucleus}  &    {}^{76}\text{Ge}	 &  {}^{82}\text{Se}  &  {}^{100}\text{Mo}  &   {}^{130}\text{Te}  &   {}^{136}\text{Xe}  &   {}^{150}\text{Nd}\!\\ 
   \hline  
\smash{\raise-1.3ex\hbox{\!\!\cite{Hirsch:1996qw}}}	&  G |\mathcal{M}_{\nu}|^2{\scriptscriptstyle\times} 10^{13}\,\text{yr} & 1.1 & 4.3 & 2.0 &  5.3 &   1.2 & 75.6 \\ 
		&      p\, /\MeV    	& 190  	& 186 &   189  &   180  &   280  &   210  \\ 
   \hline
\smash{\raise-1.3ex\hbox{\!\!\cite{Simkovic:2010ka}}} &  G |\mathcal{M}_{\nu}|^2{\scriptscriptstyle\times}10^{13}\,\text{yr}	& 2.7 & - & 15.2 &  12.2 &  - & - \\ 
		&     p\, /\MeV          &184  	&   - & 193 &   198  &  - &   - \\ 
 \hline 
\end{array} 
$$
\vspace*{-3ex}
\caption{Nuclear factors relevant for \BB.}
\label{tab:nvirtuality}
\end{table}
  
\pagebreak[3]

To illustrate the impact of the Dirac and Majorana phases on the total decay rate, we plot in the
left frame of Fig.~\ref{figTelloPlot} the well known absolute value of $m^{ee}_{\nu}$, while the
corresponding effective right-handed counterpart for the type II seesaw used here,
\begin{equation} \label{eqMNee}
	M_N^{ee} =  p^2 \frac{\displaystyle M_W^4}{\displaystyle M_{W_R}^4}  \frac{\displaystyle V_{Lej}^2}{
\displaystyle  m_{N_j}}\,,
\end{equation}
is shown 
in the right frame.  The plot was made using Eqs.~\eqref{spectrum}, \eqref{rightmixing},
 with $p = 190\,\MeV$ and taking the entire range of $V_L$ to be allowed by LFV, see
Fig.~\ref{figLFV}.

\begin{figure}[t]
	\includegraphics[height=.53\columnwidth]{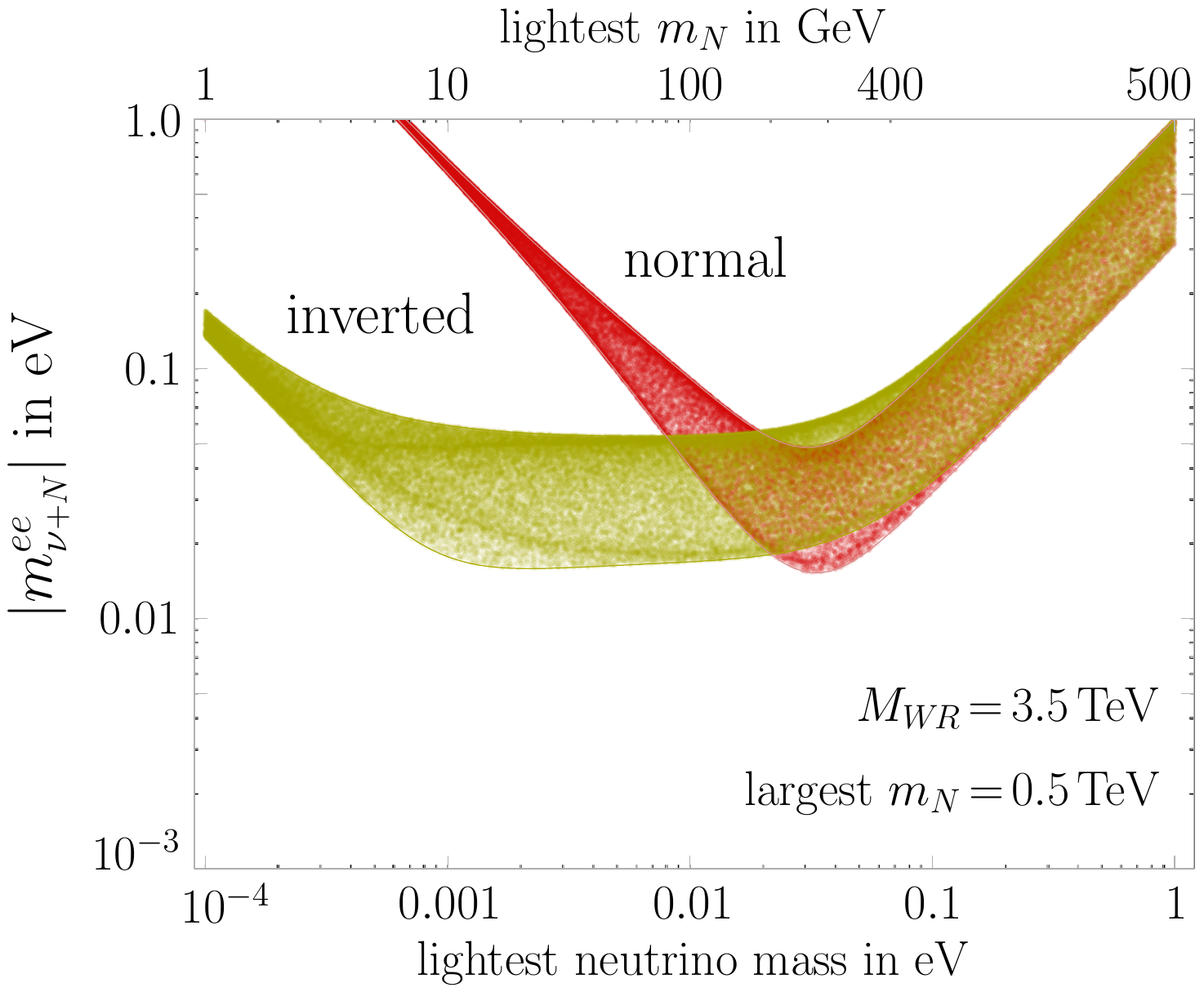}\hspace*{1em}%
       \vspace*{-1ex}%
	\caption{\label{figbetaLR}\it
	Effective \BB\ mass parameter $|m_ {\scriptscriptstyle { \nu +N } }^{ee}|$, 
	a measure of the total \BB\ rate including contributions from both left and right currents.}   
     \vspace{-0.1cm}
\end{figure}

The total \BB\ rate is governed by the effective mass parameter
\vspace*{-2ex}
\begin{equation}
|m_ {\nu + N}^{ee}|= (| m_\nu^{ee}|^2 + | M_N^ {ee}|^2)^{1/2}
\end{equation}
%
that supersedes the standard matrix element $m^{ee}_{\nu}$ in the parameter space
accessible to LHC.  In Fig.~\ref{figbetaLR}, we show $|m_ {\nu +N}^{ee}|$ as a function of the
lightest neutrino mass.  We have already stressed in the introduction the reversed role of the
neutrino mass hierarchies.  In the case of the right-handed contribution, the normal hierarchy 
prevails over the inverted in wide regions of the parameter space; 
for both
hierarchies, new physics can win over the neutrino mass as the source of \BB. Moreover,
Fig.~\ref{figbetaLR} shows that there is no more room for a vanishing transition rate, as 
in Fig.~\ref{figTelloPlot}. On the upper horizontal axis of Fig.~\ref{figbetaLR} we also
display the lightest of the heavy neutrinos. As one can see, the range of $m_N^{\text{lightest}}$ is
easily below 100\,\GeV\ which would lead to interesting displaced vertices at LHC~\cite{Maiezza:2010ic}.

In short, \BB\     
may be naturally governed by new physics and thus be disjoint from
light neutrino masses. This is only in\linebreak[3] apparent contradiction with the often stated
result~\cite{Schechter:1981bd}, according to which a non vanishing \BB\ implies a
nonvanishing neutrino Majorana mass.  Although true as a generic statement,
  on a quantitative level it has no practical purpose, as the case exposed here
demonstrates explicitly. Another example was provided by the minimal supersymmetric standard model \cite{Allanach:2009iv}.

\SEC{Discussion and outlook.} In this Letter we have shown how the minimal LR symmetric theory
offers a deep connection between high energy collider physics and low energy processes such as
neutrinoless double beta decay and lepton flavor violation. The crucial point is lepton number
violation which at LHC would reveal itself through same sign di-leptons produced from the decay of
a heavy right-handed neutrino. The different flavor channels will be a probe of the right-handed
mixing matrix, allowing to test the type-II seesaw hypothesis in the near future.

At the same time, the low scale of LR symmetry implies a sizable contribution to the neutrinoless
double beta decay rate.  The standard hypothesis that this transition is dominated by the Majorana
mass of light neutrinos may lead to a tension between oscillations and measurements of the absolute
neutrino mass.  The alternative hypothesis does not only permit wider possibilities, such as small
neutrino masses with normal hierarchy ordering and large rate for the neutrinoless double beta
decay, but much more interestingly, it has a real chance of being tested at the LHC.  


Measurements of heavy neutrinos at the LHC can easily invalidate the specific version of the model, requiring e.g., to abandon ${\mathcal C}$ symmetry and/or type II seesaw, and to replace our hypotheses on 
$m_N$ and $V_R$, Eqs.~(\ref{spectrum}) and (\ref{rightmixing}), with the experimental results. 
Whereas this would 
imply quantitative changes of our results, it would not change our main conclusion that the possible LHC findings will
be crucial for the interpretation of the neutrinoless double beta decay.

\SEC{Acknowledgments.}  We are 
grateful to A.~Melfo for 
help and encouragement throughout this work
and collaboration on related issues.  We thank B.~Bajc, S.T. Petcov 
and Y.~Zhang for useful discussions, and Y. Zhang for careful reading of the manuscript.  F.N.~and
F.V.~thank 
ICTP for 
hospitality during the initial stages of this work.

\end{document}